\documentclass{aa}
\usepackage{times,graphicx,amssymb,lscape}
\usepackage{rotating}
\usepackage{natbib}
\usepackage{txfonts}
\bibpunct{(}{)}{;}{a}{}{,}  

\begin{document}

\title{Modified $p$-modes in penumbral filaments?}

\author{D. S. Bloomfield\inst{1}
	\and
        S. K. Solanki\inst{1} 
        \and
        A. Lagg\inst{1}
        \and
        J. M. Borrero\inst{2}
        \and
        P. S. Cally\inst{3}
        }

\institute{Max-Planck-Institut f\"{u}r Sonnensystemforschung, 
	   Max-Planck-Str. 2, 37191 Katlenburg-Lindau, Germany\\
	   \email{bloomfield@mps.mpg.de}
	   \and
	   High Altitude Observatory, 3450 Mitchell Lane, Boulder, 80301 
	   Colorado, USA
	   \and
           Centre for Stellar and Planetary Astrophysics, School of 
	   Mathematical Sciences, Monash University, Victoria, 3800, 
	   Australia
	   }

\date{Received date / Accepted date}

\abstract{}
	 {The primary objective of this study is to search for and identify 
wave modes within a sunspot penumbra.}
	 {Infrared spectropolarimetric time series data are inverted using a 
model comprising two atmospheric components in each spatial pixel. Fourier 
phase difference analysis is performed on the line-of-sight velocities 
retrieved from both components to determine time delays between the velocity 
signals. In addition, the vertical separation between the signals in the two 
components is calculated from the Stokes velocity response functions.}
	 {The inversion yields two atmospheric components, one permeated by a 
nearly horizontal magnetic field, the other with a less-inclined magnetic 
field. Time delays between the oscillations in the two components in the 
frequency range $2.5-4.5$~mHz are combined with speeds of atmospheric wave 
modes to determine wave travel distances. These are compared to expected path 
lengths obtained from response functions of the observed spectral lines in the 
different atmospheric components. Fast-mode (i.e., modified $p$-mode) waves 
exhibit the best agreement with the observations when propagating toward the 
sunspot at an angle $\sim$50\degr\ to the vertical.}
	 {}


\authorrunning{D. S. Bloomfield et~al.}
\titlerunning{Modified $p$-modes in penumbral filaments?}

\maketitle 

\section{Introduction}
\label{sec:int}
Disentangling the signatures of various atmospheric waves that are supported 
by structured magnetic atmospheres is a difficult, even daunting, task. Recent 
advances in dynamic, 2-D modeling of magnetised atmospheres 
\citep{2002ApJ...564..508R, 2003ApJ...599..626B, 2006ApJ...653..739K}, 
highlight the dilemma that observers face -- i.e., measuring not just singular 
wave modes but the superposition of many magneto-atmospheric modes that most 
likely propagate in different directions and exist in differing plasma-$\beta$ 
environments. This poses a distinct problem, especially if studies do not 
resolve the true size scale of solar magnetic features. 

However, information may be extracted from spatially unresolved structures by 
spectropolarimetry \citep[e.g., the penumbral flux-tube work 
of][]{2002A&A...393..305M}. This approach uses the full Stokes polarization 
spectra ($I$, $Q$, $U$, $V$), allowing physical properties of the emitting 
plasma to be inferred through the application of appropriate model 
atmospheres. The suitability of using Stokes profiles for wave diagnostics has 
been shown through numerical simulations \citep[see, 
e.g.,][]{1992MNRAS.256...13S, 1997A&A...325.1199P, 1999A&A...345..986P}, while 
Stokes profiles were also used by \citet{1998A&A...335L..97R} to interpret 
magnetic field oscillations in a sunspot as resulting from 
magneto-acoustic-gravity waves.

In this paper we present a method that may identify the form of wave which 
exists in a magnetic environment using the information available to full 
Stokes spectropolarimetry. The observational data, their reduction, and 
details of the form of atmospheric inversion procedure applied are outlined in 
Sect.~\ref{sec:obs}. Results of response function calculations and a Fourier 
phase difference analysis are presented and discussed in Sect.~\ref{sec:res} 
in terms of the various forms of wave modes which may exist at differing 
propagation angles, while in Sect.~\ref{sec:con} the implications of our work 
are summarized.

\begin{table*}
\caption{Atomic parameters of the observed lines. $\lambda_0$ denotes the 
laboratory wavelength, $\chi_l$ the excitation potential of the lower energy 
level, and $\log gf$ the logarithm of the oscillator strength times the 
multiplicity of the level. The parameters $\alpha$ and $\sigma$ (in units of 
the Bohr radius, $a_0$) are used to calculate the line broadening by 
collisions with neutral hydrogen, atoms, while $g_l$, $g_u$, and $\bar g$ are 
the calculated Land\'{e} factors of the lower and upper levels, and the 
effective value, respectively.}
\label{tab:ato_par}
\centering
\begin{tabular}{lccccccccc}
\hline\hline
  Species	&  $\lambda_0$	&  Configuration		&  $\chi_l$	&  $\log gf$	&  $\alpha$	&  $\sigma$	&  $g_l$	&  $g_u$	&  $\bar g$\\
  		&  (\AA)	&  				&  (eV)		&  (dex)	&  		&  ($a_0$)	&  		&  		&  \\
\hline
  Fe \sc{i}	&  15662.018	&  $^{5}F_{5}-^{5}F_{4}$	&  5.83		&  0.19		&  0.24		&  1200		&  1.40		&  1.35		&  1.50\\
  Fe \sc{i}	&  15665.245	&  $^{5}F_{1}-^{5}D_{1}$	&  5.98		&  -0.42	&  0.23		&  1283		&  0.00		&  1.50		&  0.75\\
\hline
\end{tabular}
\end{table*}

\section{Observations}
\label{sec:obs}
\begin{figure}
\centering
\includegraphics[width=\columnwidth]{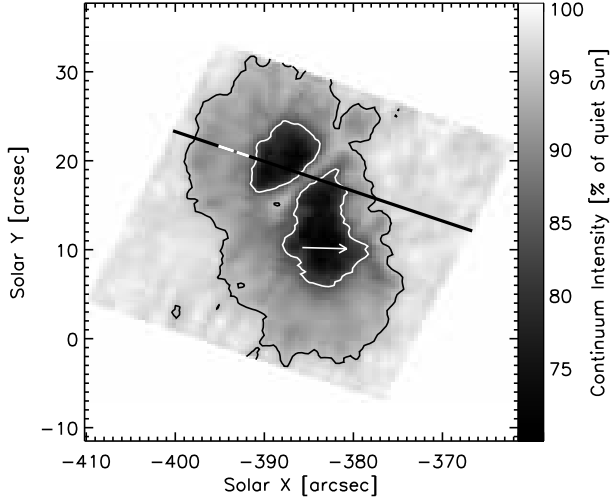}
\caption{Continuum intensity image of NOAA 10436 at 15666.5$\pm$0.5~\AA. White 
(black) contours mark the umbral/penumbral (penumbral/quiet Sun) boundaries. 
The straight black line marks the slit position during the time series and the 
white portion indicates the region studied in detail. The dark grey pixel 
within the white portion of the slit marks the position from which the 
profiles in Fig.~\ref{fig:pen_sto_pro} are taken, while the arrow in the umbra 
points to disk centre.}
\label{fig:spa_con}
\end{figure}

A small and ellipsoidal sunspot, NOAA 10436, was observed on 21 August 2003 
using the Tenerife Infrared Polarimeter \citep{1999ASPC..183..264M} attached 
to the 0.7~m German solar Vacuum Tower Telescope in Tenerife, Canary Islands. 
A fast scan was performed over the whole sunspot to obtain a global picture, 
comprising 79 slit positions of 0\farcs35 width (each with a slit integration 
time of 7~s) incrementally stepped 0\farcs4, from which the continuum 
intensity image in Fig.~\ref{fig:spa_con} was constructed. The sunspot was 
located somewhat out of disk centre ($\theta = 26.7$\degr, $\mu = \cos \theta 
\approx 0.9$) and a light bridge separated the two umbrae.

Prior to this scan, the slit was fixed across the sunspot and the full Stokes 
vector ($I$, $Q$, $U$, $V$) was measured (see Fig.~\ref{fig:pen_sto_pro} for 
example spectra) in a time series comprising 250 stationary-slit exposures 
acquired at a cadence of 14.75~s over 14:39-15:41~UT. Seeing conditions were 
moderate during the observations, with an estimated spatial resolution of 
around 1\farcs5. The most striking feature observed in the time series was the 
oscillatory behaviour of Stokes $Q$, most prominently seen in the inner part 
of the limb-side penumbra (white part of the slit in Fig.~\ref{fig:spa_con}). 
This oscillation in the Stokes $Q$ signal is diplayed in 
Figs.~\ref{fig:Q_area}{\emph{c}} and \ref{fig:Q_area}{\emph{d}}, where a 
dominant 5~min period is observed.

\subsection{Spectral Lines}
\label{subsec:spe_lin}
The recorded spectral region contains two moderately magnetically sensitive 
neutral iron lines (\ion{Fe}{i} 15662.0~\AA\ with effective Land\'{e} factor 
$\bar{g} = 1.50$ and \ion{Fe}{i} 15665.2~\AA\ with $\bar{g} = 0.75$) at a 
wavelength sampling of 29.7~m\AA~pixel$^{-1}$. The data reduction included 
polarization demodulation and calibration, flat fielding, dark current and 
continuum correction, and the removal of instrumental cross-talk between the 
Stokes profiles \citep{2003SPIE.4843...55C}. The noise in the reduced data lay 
below $5 \times 10^{-4}$ in units of continuum intensity. The wavelength 
calibration was done assuming that the core position of average quiet-Sun 
intensity profiles corresponds to the laboratory wavelength, $\lambda_0$, 
shifted towards the blue by 500~m~s$^{-1}$, the approximate value for the 
blueshift in these lines deduced from the convective velocity relation of 
\citet{1988ApJ...325..480N}.

Table~\ref{tab:ato_par} presents the atomic data for the spectral lines used 
in this work, where laboratory wavelengths, electronic configurations, and 
excitation potentials were taken from \citet{1994ApJS...94..221N} while the 
parameters $\alpha$ and $\sigma$, which are used in the calculation of 
spectral line broadening by collisions with neutral perturbers, were taken 
from \citet[][sp transitions]{1995MNRAS.276..859A}. The two-component model of 
the quiet Sun from \citet{2002A&A...385.1056B} was used to calculate empirical 
oscillator strengths for the observed lines, as in 
\citet{2003A&A...404..749B}. For the 15665~\AA\ line, the value for the 
derived oscillator strength is especially inaccurate (we estimate 
$\pm$0.2~dex) as the intensity profile is blended and the effective quantum 
number for the upper level was beyond the value given in the tables of 
\citet{1995MNRAS.276..859A}; here we take the maximum value provided by these 
authors to avoid large extrapolations.

\begin{figure}
\sidecaption
\includegraphics[width=\columnwidth]{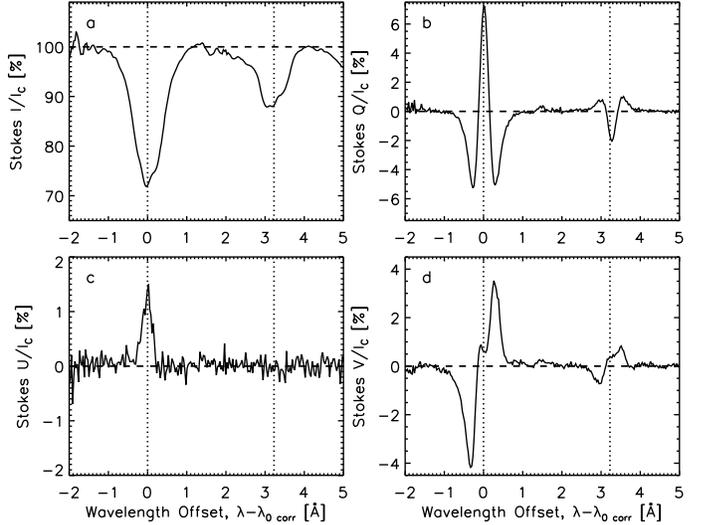}
\caption{Example penumbral profiles of Stokes $I$ ({\emph{a}}), Stokes $Q$ 
({\emph{b}}), Stokes $U$ ({\emph{c}}), and Stokes $V$ ({\emph{d}}), each 
normalized to the local continnum intensity, $I_C$, from the pixel marked dark 
grey inside the region of interest shown in Fig.~\ref{fig:spa_con}. 
Wavelengths are relative to the 15662~\AA\ line, with dotted lines at the 
blueshift-corrected laboratory values.}
\label{fig:pen_sto_pro}
\end{figure}

Figure~\ref{fig:pen_sto_pro} represents an example of the observed penumbral 
profiles, in this case from the dark grey pixel inside the region of interest 
of Fig.~\ref{fig:spa_con}. As already mentioned, the intensity profile of the 
15665~\AA\ line is blended with an unidentified profile. We identify the blend 
as solar -- it varys in strength between quiet Sun and umbra -- but its origin 
could not be determined. However, it seems that it is not magnetically 
sensitive because even when the field is strong, as in the penumbra, no 
residual polarization signal appears at that wavelength. In our analysis we 
only weakly consider the $I$ spectrum for this line but make full use of the 
polarization spectra ($Q$, $U$, $V$) because, although small in magnitude, 
they provide additional information.

The circular polarization (Stokes $V$) shows that the magnetic field points 
downward in the spot but, more interestingly, the linear polarization signals 
(Stokes $Q$ and $U$) are oppositely signed in each line. This is due to their 
opposite Zeeman patterns and can be easily proved following 
\citet{1992soti.book...71D}. In the weak magnetic field limit, it can be 
written that, 
\begin{equation}
Q \left( \lambda_0 \right) = - \frac{1}{4} \left( \bar{g}^2 + \epsilon \right) \lambda_B^2 \sin^2 \gamma \cos 2\chi \left( \frac{\partial^2 I_0}{\partial \lambda^2} \right)_{\lambda = \lambda_0} \ ,
\end{equation}
\begin{equation}
U \left( \lambda_0 \right) = - \frac{1}{4} \left( \bar{g}^2 + \epsilon \right) \lambda_B^2 \sin^2 \gamma \sin 2\chi \left( \frac{\partial^2 I_0}{\partial \lambda^2} \right)_{\lambda = \lambda_0} \ ,
\end{equation}
where $\gamma$ is the inclination of the magnetic field vector to the 
line-of-sight (LOS), $\chi$ is the azimuthal angle of the magnetic field 
vector in the plane perpendicular to the LOS, $\lambda_B$ is the Zeeman 
wavelength splitting, $\bar{g}$ is the effective Land\'{e} factor of the 
transition, and $\epsilon$ is a correction factor for anomalous Zeeman 
triplets ($J_L \neq 0$ or $J_U \neq 1$ or $g_L \neq g_U$) given by,
\begin{eqnarray}
\epsilon & = & \frac{1}{80} (  7 \left[ J_U \left( J_U + 1 \right) - J_L \left( J_L + 1 \right) \right]^2 \nonumber \\
         &   & \ \  - 16 \left[ J_U \left( J_U + 1 \right) + J_L \left( J_L + 1 \right) \right] + 4 ) \left( g_U - g_L \right)^2 \ ,
\end{eqnarray}
where $J_L$ and $J_U$ are the total angular momentum quantum numbers for the 
lower and upper transitions, respectively, as defined by the electronic 
configurations (Table~\ref{tab:ato_par}), while $g_L$ and $g_U$ are the 
Land\'{e} factors of the lower and upper levels whose expressions can be 
analytically determined as these lines are in pure L-S coupling \citep[see, 
e.g.,][]{2003isp..book.....D}. Values of $\bar{g} + \epsilon$ can be 
calculated, yielding $\simeq$2.2 for \ion{Fe}{i} 15662~\AA\ and $\simeq$$-1.1$ 
for \ion{Fe}{i} 15665~\AA, explaining the opposite linear polarization signals 
observed in Fig.~\ref{fig:pen_sto_pro}. Note, the inversion technique detailed 
in the following section makes use of the full Zeeman splitting pattern and 
does not use the effective Land\'{e} factor.

\subsection{Atmospheric Inversion}
\label{subsec:atm_inv}
\begin{figure}
\centering
\includegraphics[width=\columnwidth]{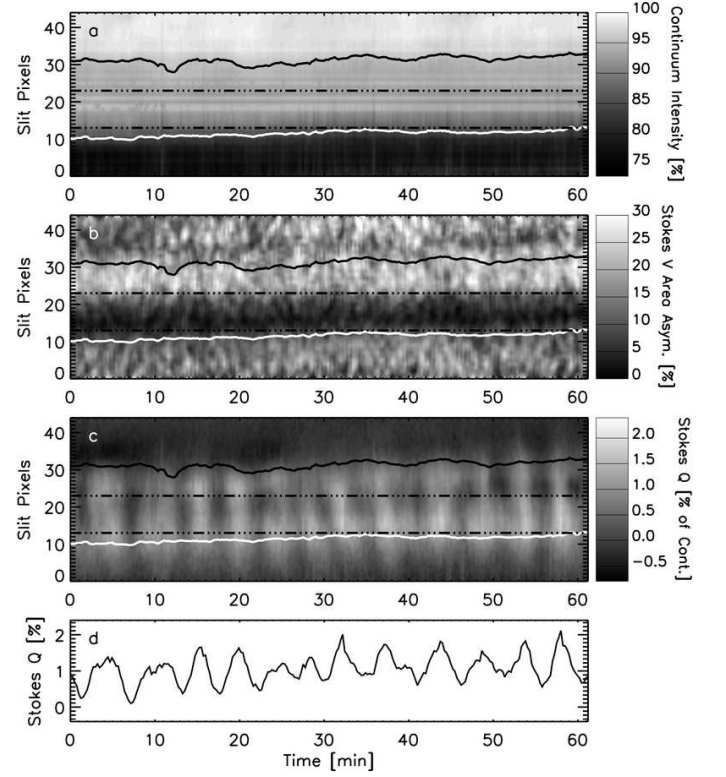}
\caption{Space-time plots of continuum intensity as a percentage of average 
quiet-Sun continnum ({\emph{a}}), magnitude of the relative Stokes $V$ area 
asymmetry ({\emph{b}}), and Stokes $Q$ at +0.145~\AA\ from the core of the 
15662~\AA\ line ({\emph{c}}). {\emph{d}}): Spatially-averaged Stokes $Q$ 
signal from the region of interest. Only the slit portion extending toward 
solar north east from the northern umbra is shown in {\emph{a}}-{\emph{c}}. 
The white (black) contour marks the umbral/penumbral (penumbral/quiet Sun) 
boundary and the dot-dashed lines bound the region studied (white area of slit 
in Fig.\ref{fig:spa_con}).}
\label{fig:Q_area}
\end{figure}

The data were inverted using the SPINOR inversion code 
\citep{2000PhDT........19F, 2000A&A...358.1109F}. Prior to the inversion we 
computed the relative Stokes $V$ area asymmetry \citep{2004A&A...422.1093B}, 
defined as,
\begin{equation}
\delta A = \frac{\int V \left( \lambda \right) d \lambda}{\int \vert V \left( \lambda \right) \vert d \lambda} \ .
\end{equation}
Given the small magnitude of $\delta A$ observed in the circular 
polarization over the inner limb-side penumbra 
(Fig.~\ref{fig:Q_area}{\emph{b}}; area between the dot-dashed lines) it is 
likely that these lines are not affected by strong gradients in either the 
magnetic field vector or velocity along the LOS. Therefore, we have adopted 
the same model as used by \citet{2004A&A...422.1093B} and referred to as a 
two-component (2-C) inversion. This model consists of two magnetic components 
and one non-magnetic straylight component -- all parameters except temperature 
are height independent in each component. The inversion returns the 
thermodynamic, magnetic, and kinematic structure of the atmosphere that 
provides the best fit to the observed Stokes ($I$, $Q$, $U$, $V$) polarization 
signals.

\begin{figure}[!hb]
\centering
\includegraphics[width=\columnwidth]{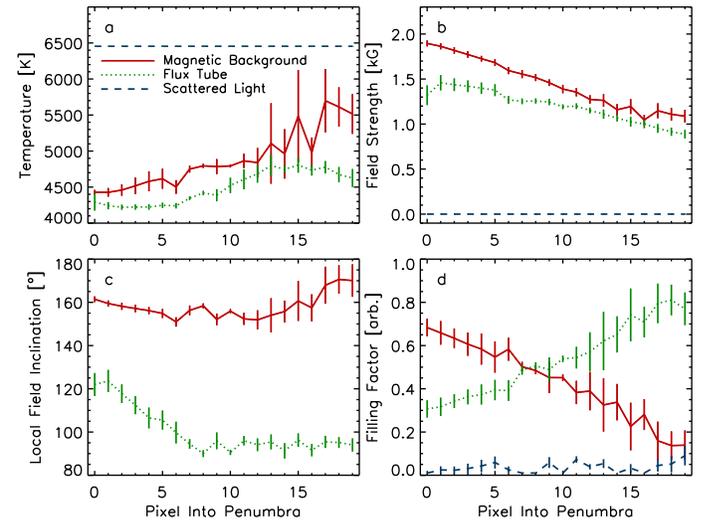}
\caption{Variation through the limb-side penumbra of the parameters obtained 
by the inversion at $\log \left( \tau \right) = 0$\,: temperature 
({\emph{a}}), field strength ({\emph{b}}), local solar inclination 
({\emph{c}}), and filling factor ({\emph{d}}). Points mark the temporal mean 
for each spatial pixel and error bars extend over $\pm$1$\sigma$.}
\label{fig:rad_var}
\end{figure}

Parameters retrieved from the inner limb-side penumbra 
(Fig.~\ref{fig:rad_var}; pixels $1-11$) yield a magnetic geometry consisting 
of a near-horizontal component, at local solar inclinations $\gamma^{\prime} 
\approx 90-125$\degr\ ($55-90$\degr\ from vertical), and a closer-to-vertical 
one, $\gamma^{\prime} \approx 150-160$\degr\ ($20-30$\degr\ from vertical), 
consistent with the observations of \citet{1993ApJ...403..780T}. Note that the 
retrieved values of $\gamma^{\prime}$, temperature, field strength, and 
filling factor are similar to those obtained by \citet{2004A&A...422.1093B}. 
From this point on, the near-horizontal component will be referred to as flux 
tube (FT) and the less-inclined one as magnetic background (MB) following the 
flux tube interpretation of \citet{1993A&A...275..283S}, 
\citet{1998A&A...337..897S}, \citet{2002A&A...393..305M}, and 
\citet{2005A&A...436..333B, 2006A&A...450..383B}. Note that this 
interpretation is subject to current discussion \citep{2006A&A...447..343S}. 
In principle, it may be possible to distinguish between the two scenarios 
by means of an analysis similar to ours, but this requires further development 
of the Spruit and Scharmer scenario and is beyond the scope of the current 
paper.

\section{Results \& Discussion}
\label{sec:res}

\subsection{Height Separations}
\label{subsec:hgt_sep}
\begin{figure}
\centering
\includegraphics[width=\columnwidth]{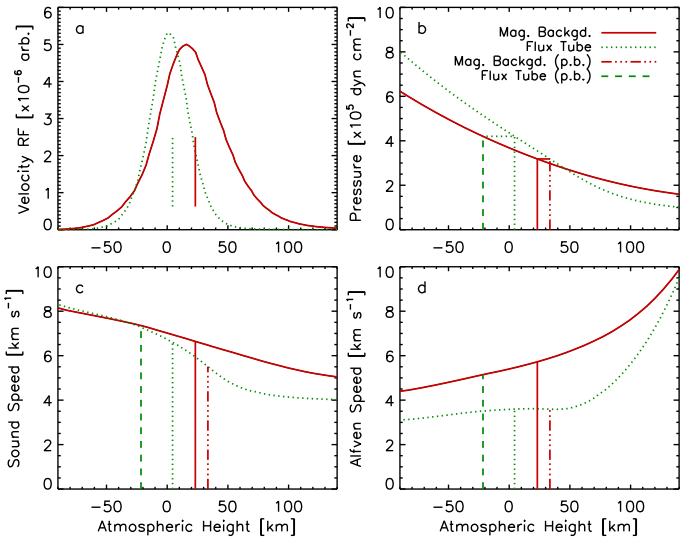}
\caption{Height variation of wavelength-integrated 15662~\AA\ and 15665~\AA\ 
Stokes $Q$ velocity response ({\emph{a}}), total pressure ({\emph{b}}), 
acoustic ({\emph{c}}), and Alfv\'{e}nic ({\emph{d}}) wave speeds for the 
magnetic background and flux tube atmospheres (solid and dotted curves, 
respectively). Vertical lines in {\emph{a}} mark the COG in each component; 
vertical dashed and dot-dashed lines in {\emph{b}}-{\emph{d}} show these 
heights translated into the reference frame of the other atmosphere by 
enforcing total pressure balance.}
\label{fig:hei_pre_bal}
\end{figure}

In the following analysis the height at which the oscillations are observed in 
both components plays an important role, so that we describe in detail how 
this height is determined. We compute the line depression response function 
(RF), which is most appropriate for the Stokes parameters \citep{1988A&A...204..266G}. The 
wavelength-integrated RFs of Stokes $Q$ to LOS velocity are presented in 
Fig.~\ref{fig:hei_pre_bal}{\emph{a}} since the velocity variations were most 
strongly observed in Stokes $Q$ (Fig.~\ref{fig:Q_area}). The RFs for each of 
the components overlap over most of the atmosphere, but their 
center-of-gravity (COG), displayed as vertical lines, reveal a distinct 
separation between the components. These COG heights were taken as the origin 
of the LOS velocity signals.

One problem with comparing these COG heights is that each refers to the height 
scale of the atmosphere in which the lines were computed. The atmospheres 
returned by the inversion each have their own height scale due to the different 
effective temperatures (Fig.~\ref{fig:rad_var}{\emph{a}}) yielding different 
pressure scale heights (Fig.~\ref{fig:hei_pre_bal}{\emph{b}}). As such, they 
cannot be simply reduced to a single height scale, since pressure balance 
between the two atmospheres can only be enforced at a single height at a time.

In order to determine vertical height separations between the COG heights of 
the two components this height scale inequality must be overcome. This was 
achieved by enforcing total pressure balance between the two atmospheres at 
the COG heights. For example, in Fig.~\ref{fig:hei_pre_bal}{\emph{b}} the 
value from the FT pressure curve at the FT COG (dotted vertical line) is 
translated onto the MB pressure curve, yielding its pressure-balanced COG in 
the MB reference frame (dashed line). Similarly, the value from the MB 
pressure curve at the MB COG (solid vertical line) is translated onto the FT 
pressure curve, providing its pressure-balanced COG in the FT reference frame 
(dot-dashed line).

Through this approach we arrive at representative heights for the velocity 
signals of the MB and FT atmospheres in the reference frames of either 
atmosphere. Values are determined for both cases as a consistency check for 
the method. This allows the vertical separation distances of the velocity 
signals to be calculated in either of the reference frames -- in the MB (FT) 
atmosphere it is the distance between the solid and dashed (dotted and 
dot-dashed) vertical lines. These heights also allow the retrieval of 
characteristic wave propagation speeds from the output inversion atmospheres 
(e.g., Figs.~\ref{fig:hei_pre_bal}{\emph{c}} and 
\ref{fig:hei_pre_bal}{\emph{d}}).

\subsection{Time Series Analysis}
\label{subsec:tim_ser}
All atmospheric parameters except velocity were found to be time independent 
within the error bars. Clear periodic LOS velocity variations were observed in 
both atmospheric components, prompting the application of Fourier phase 
difference analysis following \citet{2001A&A...379.1052K}. The analysis was performed 
between the MB and FT velocity signals of eleven pixels of the inner limb-side 
penumbra. The phase difference from one pixel is the frequency-dependent phase 
lag of the FT velocity signal with respect to the MB signal, while the 
coherence measures the quality of phase difference variation.

\begin{figure}
\sidecaption
\includegraphics[width=\columnwidth]{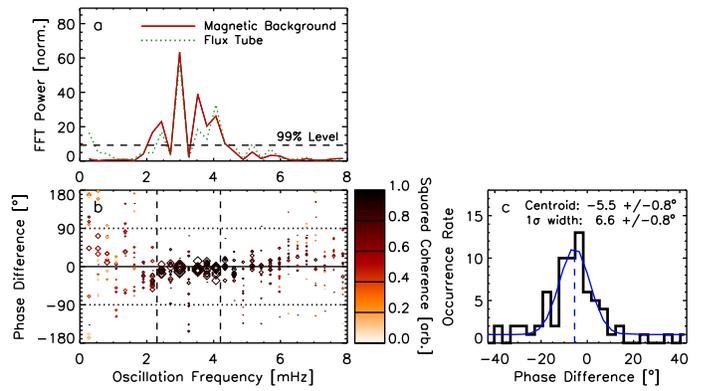}
\caption{{\emph{a}}): Fourier power spectra from co-spatial magnetic 
background and flux tube atmosphere LOS velocities (solid and dotted curves, 
respectively). {\emph{b}}): Fourier phase difference spectra between the 
magnetic background and flux tube velocities from the eleven analyzed pixels. 
Darker shading denotes greater Fourier coherence and larger symbol size 
greater cross-spectral power. {\emph{c}}): PDF of phase difference values over 
the range $2.5-4.5$~mHz. The thick curve displays the measured values and the 
thin curve the best-fit Gaussian profile to the data.}
\label{fig:fft_pha_dif}
\end{figure}

Figure~\ref{fig:fft_pha_dif} displays the output from applying such 
an analysis to the observed velocities. Example Fourier power spectra from one 
spatial pixel are given in Fig.~\ref{fig:fft_pha_dif}{\emph{a}} where both 
components obviously show power at very similar frequencies. The overplotted 
phase difference spectra for all eleven of the analyzed spatial pixels is 
presented in Fig.~\ref{fig:fft_pha_dif}{\emph{b}}, where symbol size 
represents the magnitude of cross-spectral power and shading denotes the 
coherence. Phase difference values are approximately constant in the range 
showing strongest cross-spectral power ($\sim$$2.5-4.5$~mHz) and, although 
close to zero, the probability distribution function (PDF) in 
Fig.~\ref{fig:fft_pha_dif}{\emph{c}} reveals that they are centred on 
$-5.5$\degr. This centroid phase difference value was converted to a time 
delay between the signals, resulting in values of $-6.3$~s to $-3.8$~s over 
the detected range of oscillation frequencies. Negative phase differences mean 
that the FT velocity leads the MB, agreeing with the COG heights in 
Fig.~\ref{fig:hei_pre_bal} for upward wave propagation.

\subsection{Wave Modes}
\label{subsec:wav_mod}
The range of atmospheric parameters existing in a sunspot penumbra supports 
the possibility of many differing forms of wave modes. However, all of these 
could be excited by $p$-mode waves propagating up towards the solar surface 
and the fact that the power peaks at $\sim$3~mHz 
(Fig.~\ref{fig:fft_pha_dif}{\emph{a}}) suggests that this is indeed the case 
for the observed wave modes. The dispersion relation for acoustic (i.e., 
$p$-mode) waves has the form $k^2 c_S^2 = \omega^2 - \omega_{\rm{ac}}^2$, 
where $c_S$ is the sound speed, $\omega$ is the angular frequency, and 
$\omega_{\rm{ac}}$ is the acoustic cutoff frequency: waves are evanescent for 
$k^2 < 0$ and propagation occurs for $\omega > \omega_{\rm{ac}}$.

\begin{figure}
\centering
\includegraphics[width=8.0cm]{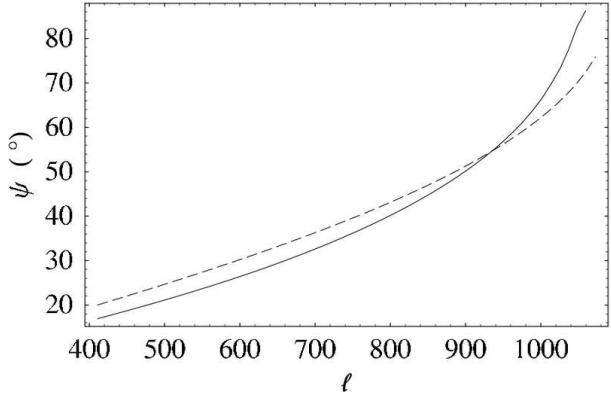}
\caption{Variation of propagation angle from the vertical, $\psi$, with 
angular wave mode, $l$, for acoustic waves at a frequency of $3.5$~mHz. The 
solid curve shows the case 50~km below the equipartition level (where the 
acoustic and Alfv\'{e}n speeds are equal) in a 1.75~kG field inclined at 
20\degr, while the dashed curve is 90~km below the equipartition level in a 
1.5~kG field inclined at 75\degr\ (the sampling heights and configurations of 
the magnetic background and flux tube components, respectively).}
\label{fig:pmod_inc_ang}
\end{figure}

\begin{table*}
\caption{Differences between COG values of RF-predicted and calculated path 
length distributions. Values are presented for cases of waves excited by 
$p$-modes with inclinations, $\psi$, of 0\degr, 40\degr, 50\degr, and 60\degr\ 
from the vertical. Note that in the case of vertical propagation 
($\psi=0$\degr) there is no distiction between ingressing or egressing forms 
of fast-mode waves, hence only one value is provided. When sampling a low 
plasma-$\beta$ environment the Alfv\'{e}n and slow-mode waves are field 
aligned, remaining invariant to changes in the originating $p$-mode 
inclination: values are not provided for these two wave modes in the flux-tube 
reference frame due to their large magnitude and obvious incorrectness.}
\label{tab:cog_dif_val}
\centering
\begin{tabular}{lcccccccc}
\hline\hline
  Wave Mode			&  \multicolumn{8}{c}{Absolute Separation of COG Values [km]}\\
  				&  \multicolumn{2}{c}{$\psi=0$\degr}	&  \multicolumn{2}{c}{$\psi=40$\degr}	&  \multicolumn{2}{c}{$\psi=50$\degr}	&  \multicolumn{2}{c}{$\psi=60$\degr}\\
  				&  MB		&  FT		&  MB	&  FT		&  MB	&  FT		&  MB	&  FT\\
\hline
  Alfv\'{e}nic			&  5		&  \ldots	&  5	&  \ldots	&  5	&  \ldots	&  5	&  \ldots\\
  Slow				&  9		&  \ldots	&  9	&  \ldots	&  9	&  \ldots	&  9	&  \ldots\\
  Acoustic			&  7		&  12		&  2	&  6		&  8	&  1		&  20	&  7\\
  Fast (ingressing)		&  10		&  16		&  5	&  9		&  0	&  4		&  12	&  5\\
  Fast (egressing)		&  \ldots	&  \ldots	&  1	&  8		&  5	&  2		&  16	&  6\\
\hline
\end{tabular}
\end{table*}

Attributing the observed phase differences in the $2.5-4.5$~mHz range to 
propagating waves conflicts with this simple picture of evanescent behaviour 
as the cutoff frequency at the photosphere is $\sim$5~mHz in the isothermal 
case. However, in the presence of a magnetic field the acoustic cutoff is 
reduced for non-vertical waves in a rather complicated manner 
\citep{1977A&A....55..239B}. The largest deviation occurs for waves in the 
strong-field limit (i.e., when the Alfv\'{e}n speed, $v_A$, is much greater 
than $c_S$). In this situation the cutoff frequency is lowered to 
$\omega_{\rm{ac}} \cos \gamma^{\prime}$ -- termed the ramp effect -- where 
$\gamma^{\prime}$ is the magnetic field inclination from the vertical. 
Furthermore, $p$-modes may travel at angles away from the vertical at the 
heights sampled here. This is illustrated in Fig.~\ref{fig:pmod_inc_ang} using 
ray-theory calculations like those of \citet{2007AN....328..286C} for 3.5~mHz 
waves in environments similar to our two magnetic components: angles of $40 - 
60$\degr\ are possible for angular modes $l \approx 760 - 980$ around the 
sampling heights of the MB and FT components.

The situation is further complicated here as these data are not recorded in 
the strong field limit \citep[Fig.~\ref{fig:hei_pre_bal}; $c_S \geqslant v_A$ 
below 70~km and 90~km in the MB and FT atmospheres, respectively; previously 
noted by][]{1993A&A...277..639S} and the waves are influenced by two separate 
magnetic inclinations. If we assume that incident $p$-mode waves actually 
experience some average between the differing magnetic inclinations of the MB 
and FT components then the effective value of field inclination will be in the 
range $40 - 60$\degr. As such, investigation of Fig.~1 in 
\citet{1977A&A....55..239B} yields an expected reduction of the cutoff 
frequency to $\sim$$0.8\,\omega_{\rm{ac}}$ ($\approx$4~mHz) for the case where 
$c_S \approx v_A$, with an absolute maximum reduction to 
$0.5\,\omega_{\rm{ac}}$ ($\approx$2.5~mHz) in the strong field limit. However, 
we note that the concept of a cutoff frequency is somewhat questionable 
\citep[see discussions in][]{2006MNRAS.372..551S, 2007AN....328..286C}, 
especially its discrete nature if radiative cooling is considered 
\citep{1980SoPh...68...87W}.

The low-photospheric sampling of the spectral lines means that both components 
are mostly gas dominated over the heights sampled 
(Fig.~\ref{fig:hei_pre_bal}). As mentioned previously, differing forms of wave 
can exist, each having certain properties in terms of their propagation speed 
and direction: isotropic acoustic waves propagate at $c_S$; Alfv\'{e}n waves 
are restricted to the direction of the field and move at $v_A$; 
magneto-acoustic slow modes propagate along the field at $v_A$ or, if the 
sunspot is considered a large ``flux tube'', the tube speed, $c_T^2 = c_S^2 
v_A^2 / \left( c_S^2 + v_A^2 \right)$; magneto-acoustic fast modes move at,
\begin{equation}
v_F^2 = \frac{ c_S^2 + v_A^2 }{2} + \frac{ \sqrt{ \left( c_S^2 + v_A^2 \right)^2 - 4 c_S^2 v_A^2 \left( \cos \alpha \right)^2 } }{ 2 } \ ,
\end{equation}
where $\alpha$ is the angle between the direction of the field 
($\gamma^{\prime}$) and that of wave propagation ($\psi$). Note that the 
fast-mode speed differs depending on the initial direction of the $p$-mode 
waves exciting these wave modes, the most extreme difference being between the 
case of waves moving toward the sunspot (ingressing; $\alpha = \psi + 
\gamma^{\prime}$) and those moving away (egressing; $\alpha = \psi - 
\gamma^{\prime}$) when the wave vector occurs in the same azimuthal plane as 
that of the magnetic field.

\begin{figure}
\centering
\includegraphics[width=\columnwidth]{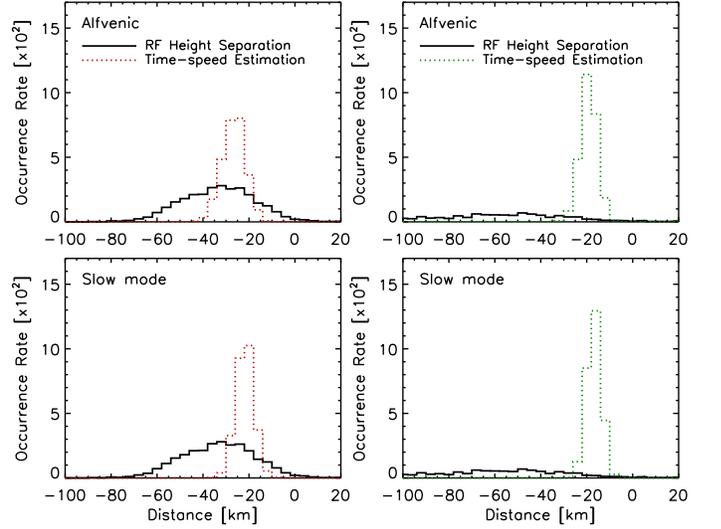}
\caption{Comparison of RF-predicted (solid lines) and calculated (dotted 
lines) wave travel distances in the reference frame of the magnetic background 
({\emph{left}}) and the flux tube ({\emph{right}}). Cases are presented 
for field-aligned waves propagating at the Alfv\'{e}n ({\emph{top}}) and 
slow-mode tube speeds ({\emph{bottom}}). The RF-predicted vertical height 
separations have been converted into path length along the propagation 
direction.}
\label{fig:b_ali_hei_vel_com}
\end{figure}

\begin{figure}
\centering
\includegraphics[width=\columnwidth]{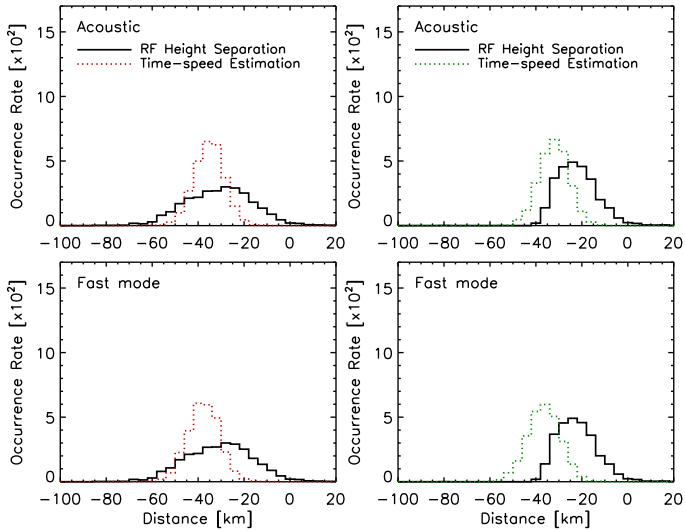}
\caption{As for Fig.~\ref{fig:b_ali_hei_vel_com}, but for acoustic 
({\emph{top}}) and fast-mode waves ({\emph{bottom}}) propagating vertically.}
\label{fig:iso_hei_vel_com_0}
\end{figure}

\begin{figure}
\centering
\includegraphics[width=\columnwidth]{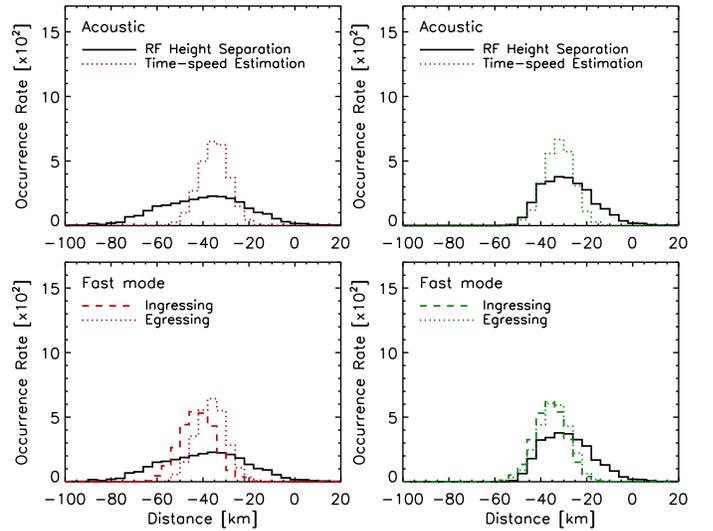}
\caption{As for Fig.~\ref{fig:b_ali_hei_vel_com}, but for acoustic 
({\emph{top}}) and fast-mode waves ({\emph{bottom}}) propagating at 
40\degr\ to the vertical.}
\label{fig:iso_hei_vel_com_40}
\end{figure}

\begin{figure}
\centering
\includegraphics[width=\columnwidth]{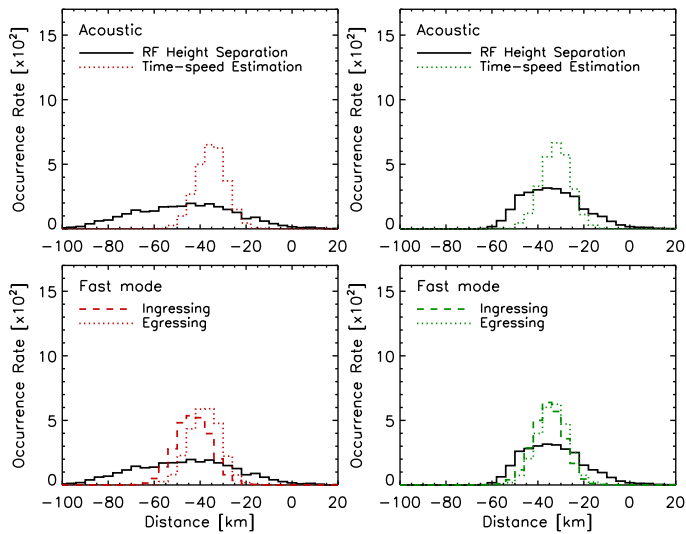}
\caption{As for Fig.~\ref{fig:b_ali_hei_vel_com}, but for acoustic 
({\emph{top}}) and fast-mode waves ({\emph{bottom}}) propagating at 
50\degr\ to the vertical.}
\label{fig:iso_hei_vel_com_50}
\end{figure}

\begin{figure}
\centering
\includegraphics[width=\columnwidth]{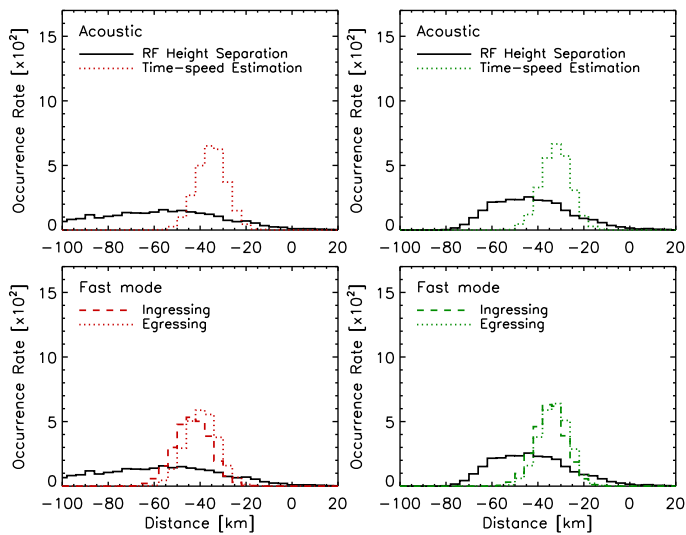}
\caption{As for Fig.~\ref{fig:b_ali_hei_vel_com}, but for acoustic 
({\emph{top}}) and fast-mode waves ({\emph{bottom}}) propagating at 
60\degr\ to the vertical.}
\label{fig:iso_hei_vel_com_60}
\end{figure}

Taking these considerations into account, the vertical height separations 
obtained in Sect.~\ref{subsec:hgt_sep} are converted into path lengths along 
the direction of propagation. Probability distribution functions of 
RF-predicted path lengths from every space-time pixel are presented in 
Figs.~\ref{fig:b_ali_hei_vel_com} to \ref{fig:iso_hei_vel_com_60} as solid 
curves, with values calculated by combining time delays, wave speeds and 
propagation angle to the field given as dotted (and in the case of 
non-vertical fast-mode waves also dashed) curves. The difference between COG 
values of the predicted and calculated distributions are given in 
Table~\ref{tab:cog_dif_val} as a quantitative measure of the correspondence 
between the various PDFs.

A comparison between the RF-predicted and calculated path lengths of 
Alfv\'{e}nic and slow-mode waves is presented in 
Fig.~\ref{fig:b_ali_hei_vel_com}, where better correspondence is observed for 
the case of Alfv\'{e}nic waves over that of slow modes in the MB reference 
frame. In the FT reference frame, however, effectively no correspondence is 
observed between the predicted and calculated PDFs due to the large values of 
field inclination. The PDF comparison for these two wave modes does not change 
for the differing values of originating $p$-mode propagation angle shown in 
Table~\ref{tab:cog_dif_val} as the Alfv\'{e}nic and slow modes remain directed 
along the magnetic field. The isotropic nature of acoustic and fast-mode waves 
expected in the sampled region of the atmosphere means that almost any angle 
of propagation could be adopted. An initial consideration of vertical 
propagation is provided in Fig.~\ref{fig:iso_hei_vel_com_0}, which shows that 
acoustic waves yield a slightly better correspondence over fast modes, 
although neither can be considered successful.

However, more appropriate values of 40\degr, 50\degr, and 60\degr\ from the 
vertical are presented based on the previous discussion of $p$-mode behaviour. 
The case of propagation at 40\degr\ to the vertical is shown in 
Fig.~\ref{fig:iso_hei_vel_com_40}, where it is unclear if acoustic or 
fast-mode waves yield better overlap between predicted and calculated PDFs. 
The distributions given in Fig.~\ref{fig:iso_hei_vel_com_50} are arrived at 
when considering propagation at 50\degr\ to the vertical. In this scenario 
fast-mode waves appear to give better correspondence than acoustic waves in 
both the MB and FT reference frames, with the case of ingressing fast modes 
providing an improvement over that of egressing fast modes. In the final case 
studied, for waves propagating at 60\degr\ to the vertical, all of the PDFs in 
Fig.~\ref{fig:iso_hei_vel_com_60} show very poor correspondence when compared 
to the cases of 40\degr\ and 50\degr\ presented above.

\section{Conclusions}
\label{sec:con}
Periodic LOS velocities retrieved from two atmospheric components in a sunspot 
penumbra have been studied to identify the form of wave mode present. The best 
correspondence between RF-predicted and calculated path lengths is observed 
for fast-mode waves propagating toward the sunspot (i.e., ingressing) at 
$\sim$50\degr\ to the vertical: this scenario has the smallest combination of 
differences between the COG separations of predicted and calculated 
distributions in the two reference frames (Table~\ref{tab:cog_dif_val}). The 
case for fast-mode ingression over egression is supported by the location of 
the sunspot being $\sim$27\degr\ from disk centre: egressing waves in the 
limb-side (and hence ingressing waves in the centre-side) penumbra will have a 
considerable component of their plasma motions directed perpendicular to our 
LOS and thus be difficult to observe; ingressing waves in the limb-side (and 
egressing waves in the centre-side) penumbra will be predominantly along our 
LOS. Further support is given by the horizontal wavelengths ($l \approx 870$, 
$\lambda \approx 5$~Mm or 7\arcsec) of the $p$-modes responsible for their 
excitation in the limb-side penumbra: egressing waves will have traversed two 
or three ``skips'' through the sunspot, including regions beneath either one 
or both of the umbrae, increasing the likelihood of their absorption or 
disruption; ingressing waves will be on their first ``skip'' into the outer 
region beneath the spot, remaining relatively strong and coherent. Despite the 
centre-side penumbra lacking strong Stokes $Q$ variations like those seen in 
the limb-side (because of the differing magnetic geometries with respect to 
the LOS), less clearly periodic velocities of $\sim$5-min period are seen 
there -- compatible with ingressing limb-side waves being disrupted after 
traversing the spot. Although these data may not fully validate results from 
local helioseismology, in regards to wave ingression and egression, this may 
be possible in the future using 2-D spectropolarimetric data -- e.g., using 
TESOS \citep{1998A&A...340..569K} in VIP mode -- allowing direct comparison 
between time-distance analysis and this technique.

This is the first time that spectropolarimetric data have been used in this 
manner to identify a magneto-acoustic wave mode. The fact that a fast-mode 
wave best fits the observational data makes qualitative sense as the spectral 
lines used here sample a high-$\beta$ region of the deep photosphere where 
$p$-mode waves are expected to be modified into fast-mode waves by the 
presence of a magnetic field. As such, this further highlights the role that 
solar internal acoustic waves may play in dynamic phenomena both at and above 
the solar surface. It will be interesting to see if the detected form of 
magneto-acoustic wave changes from the case where the velocity response of a 
spectral line is formed below the $c_S = v_A$ (i.e., $\beta \approx 1$) 
surface to one where it is formed above this level. In particular, the ratio 
of wave amplitude, or energy content, observed both above and below the $c_S = 
v_A$ level may help confirm the direction of the incident $p$-mode waves 
\citep[c.f.,][]{2006MNRAS.372..551S}. 

This paper illustrates the capability of Stokes spectropolarimetry for 
improved wave-mode identification over imaging studies, which require an 
assumption about the production of intensity variations as well as inferrence 
of the magnetic field geometry that are usually, if at all, provided by 
potential field extrapolations from LOS magnetograms. The benefits of the 
combined determination of plasma velocities and retrieval of the full magnetic 
vector appear to outweigh the reduction in spatial coverage caused by using a 
slit-based instrument. 

Finally, we note that the interpretation may depend to some extent on the 
model employed when carrying out the inversions. We have restricted ourselves 
to the simplest such two-component model. The use of more sophisticated models 
could lead to further refinements in the results.

\begin{acknowledgements}
The German solar Vacuum Tower Telescope is operated on Tenerife by the 
Kiepenheuer Insitute in the Spanish Observatorio del Teide of the Instituto de 
Astrof\'{i}sica de Canarias.
\end{acknowledgements}


\end{document}